\documentclass[iop]{emulateapj}
\usepackage{amsmath}

\newcommand{\Kalpha}{K$\alpha$}

\newcommand{\logT}{\ensuremath{\log(T/\mathrm{K})}}

\newcommand{\Msol}{\ensuremath{M_{\odot}}}
\newcommand{\Ne}{\ensuremath{n_{\mathrm{e}}}}

\newcommand{\nm}{\ensuremath{\mbox{\nm}}}
\newcommand{\cm}{\ensuremath{\mbox{cm}}}

\newcommand{\km}{\ensuremath{\mbox{km}}}

\newcommand{\pc}{\ensuremath{\mbox{pc}}}
\newcommand{\kpc}{\ensuremath{\mbox{kpc}}}

\newcommand{\s}{\ensuremath{\mbox{s}}}
\newcommand{\ks}{\ensuremath{\mbox{ks}}}

\newcommand{\Myr}{\ensuremath{\mbox{Myr}}}

\newcommand{\kev}{\ensuremath{\mbox{keV}}}

\newcommand{\erg}{\ensuremath{\mbox{erg}}}
\newcommand{\microgauss}{\ensuremath{\mu\mbox{G}}}

\newcommand{\K}{\ensuremath{\mbox{K}}}

\newcommand{\pdegsq}{\ensuremath{\mbox{deg}^{-2}}}
\newcommand{\pcc}{\ensuremath{\cm^{-3}}}
\newcommand{\pcmsq}{\ensuremath{\cm^{-2}}}

\newcommand{\ps}{\ensuremath{\s^{-1}}}

\newcommand{\emismeas}{\ensuremath{\cm^{-6}}\ \pc}
\newcommand{\ergps}{\erg\ \ps}
\newcommand{\flux}{\erg\ \pcmsq\ \ps}

\newcommand{\kmps}{\km\ \ps}

\newcommand{\chandra}{\textit{Chandra}}

\newcommand{\rosat}{\textit{ROSAT}}

\newcommand{\xmm}{\textit{XMM-Newton}}

\newcommand{\citepossessive}[1]{\citeauthor{#1}'s \citeyearpar{#1}}

\providecommand{\eqref}[1]{equation~(\ref{#1})}

\newcommand{\hs}{HS13}
\newcommand{\hskjm}{H10}
\newcommand{\hill}{H12}
\newcommand{\jm}{JM06}
\newcommand{\Stotal}{\ensuremath{S_{0.5-2.0}}}
\newcommand{\Th}{\ensuremath{T_\mathrm{h}}}
\newcommand{\Tc}{\ensuremath{T_\mathrm{c}}}
\newcommand{\nc}{\ensuremath{n_\mathrm{c}}}
\newcommand{\neh}{\ensuremath{n_\mathrm{e,h}}}
\newcommand{\nmean}{\ensuremath{\langle\Ne\rangle}}
\newcommand{\nms}{\ensuremath{\langle\Ne^2\rangle}}
\newcommand{\nrms}{\ensuremath{\nms^{1/2}}}
\newcommand{\rms}{r.m.s.}
\newcommand{\bx}[1]{\texttt{bx#1}}
\newcommand{\zmax}{\ensuremath{z_\mathrm{max}}}
\newcommand{\LCX}{\ensuremath{L_\mathrm{CX}}}

\shorttitle{TESTING GALACTIC FOUNTAIN MODELS' X-RAY EMISSION}
\shortauthors{HENLEY ET AL.}

\begin{document}

\title{The Origin of the Hot Gas in the Galactic Halo: Testing Galactic Fountain Models' X-ray Emission}
\author{David B. Henley\altaffilmark{1},
        Robin L. Shelton\altaffilmark{1},
        Kyujin Kwak\altaffilmark{2},
        Alex S. Hill\altaffilmark{3,5}, and
        Mordecai-Mark Mac Low\altaffilmark{4}
}
\affil{$^1$Department of Physics and Astronomy, University of Georgia, Athens, GA 30602, USA; dbh@physast.uga.edu \\
       $^2$School of Natural Science, Ulsan National Institute of Science and Technology (UNIST), \\
           50 UNIST-gil, Ulju-gun, Ulsan, Republic of Korea, 689-798 \\
       $^3$CSIRO Astronomy \& Space Science, Marsfield, NSW, Australia \\
       $^4$Department of Astrophysics, American Museum of Natural History, 79th Street at Central Park West, New York, NY 10024, USA
}
\altaffiltext{5}{Current address: Department of Astronomy, Haverford College, 370 Lancaster Avenue, Haverford, PA 19041, USA}

\begin{abstract}
  We test the X-ray emission predictions of galactic fountain models against \xmm\ measurements of
  the emission from the Milky Way's hot halo. These measurements are from 110 sight lines, spanning
  the full range of Galactic longitudes.
  We find that a magnetohydrodynamical simulation of a supernova-driven interstellar medium, which
  features a flow of hot gas from the disk to the halo, reproduces the temperature but significantly
  underpredicts the 0.5--2.0~\kev\ surface brightness of the halo (by two orders of magnitude, if we
  compare the median predicted and observed values). This is true for versions of the model with and
  without an interstellar magnetic field.
  We consider different reasons for the discrepancy between the model predictions and the
  observations. We find taking into account overionization in cooled halo plasma, which could in
  principle boost the predicted X-ray emission, is unlikely in practice to bring the predictions in
  line with the observations. We also find that including thermal conduction, which would tend to
  increase the surface brightnesses of interfaces between hot and cold gas, would not overcome the
  surface brightness shortfall. However, charge exchange emission from such interfaces, not included
  in the current model, may be significant.
  The faintness of the model may also be due to the lack of cosmic ray driving, meaning that the
  model may underestimate the amount of material transported from the disk to halo. In addition, an
  extended hot halo of accreted material may be important, by supplying hot electrons that could
  boost the emission of the material driven out from the disk. Additional model predictions are
  needed to test the relative importance of these processes in explaining the observed halo
  emission.
\end{abstract}


\section{INTRODUCTION}
\label{sec:Introduction}

\setcounter{footnote}{5}

X-ray observations show that the halo of our galaxy contains hot, diffuse plasma. This plasma is
observed both in emission, as a component of the $\sim$0.1--1~\kev\ soft X-ray background
\citep[SXRB; e.g.,][]{kuntz00,yoshino09,henley10b,henley13}, and in absorption, in high-resolution
X-ray spectra of bright background sources
\citep{nicastro02,rasmussen03,mckernan04,fang06,bregman07,yao07a,yao09,hagihara10,gupta12}.
Although the Milky Way's hot halo is well studied observationally, the details of the origin of this
hot plasma remain uncertain. Understanding the relative importance to the hot halo of supernova (SN)
driven outflows from the disk and inflows from the intergalactic medium is a key part of
understanding the functioning of the Galaxy and its interaction with its environment.

In \citet[hereafter \hskjm]{henley10b}, we tested models of the hot halo plasma against 26
\xmm\ observations of the high-latitude SXRB, by comparing the observed temperatures and emission
measures of the halo with the distributions predicted by different physical models. \hskjm's
analysis favored SN-driven galactic fountains \citep[hereafter \jm]{joung06} as a major, possibly
dominant, source of the hot halo plasma observed in emission, although these fountain models tended
to overpredict the halo temperature. Additional support for the heating of the halo by disk SNe
comes from the observation that some models of the halo's global gas distribution (constrained by
various observational data) imply that the hot halo may be convectively unstable
\citep{henley14a}. However, \hskjm\ were unable to rule out the possibility that an extended halo of
accreted material also contributes to the emission \citep{crain10}.

In this paper, we further examine the X-ray predictions of galactic fountain models, in the light of
two developments since \hskjm. First, we use a much larger set of measurements of the Galactic halo
emission. \citet[hereafter \hs]{henley13} measured the halo X-ray emission on 110 high-latitude
\xmm\ sight lines, an approximately fourfold increase over \hskjm. This is the largest set of
measurements of the halo X-ray emission with CCD-resolution spectra assembled to date.  Furthermore,
these observations span the full range of Galactic longitudes, whereas \hskjm's data set was
restricted to $l = 120\degr$--240\degr. See Section~\ref{sec:Observations} for a description of the
observational data. (Note that \hs\ discussed the energetics of galactic outflows versus
extragalactic accretion as sources of the observed X-ray emission, but were unable to distinguish
between these two scenarios: both SNe and infall provide more than enough energy to power
the observed emission, and either process could plausibly explain the observed variation of the
surface brightness on the sky.)

Second, it has been discovered that the \jm\ fountain model included an unphysical inflow of hot gas
from the vertical boundaries that adversely affected the model's X-ray predictions (see
Section~\ref{sec:FountainModel}). We therefore examine a new model of the SN-driven interstellar
medium (ISM), in which there is no such hot inflow \citep[hereafter \hill]{hill12}. In addition,
these newer simulations include results obtained with a non-zero interstellar magnetic field. See
Section~\ref{sec:FountainModel} for a description of these ISM models.

Section~\ref{sec:FountainCharacterization} describes how we obtained the X-ray predictions from the
ISM models. We present the results of the comparison of the model predictions with the observations
in Section~\ref{sec:Results}. We discuss the results in Section~\ref{sec:Discussion}, and conclude
with a summary in Section~\ref{sec:Summary}.

\section{OBSERVATIONS}
\label{sec:Observations}

We use \hs's measurements of the Galactic halo X-ray emission.  They measured this emission on 110
high-latitude \xmm\ sight lines, selected from an all-sky \xmm\ survey of the SXRB
\citep{henley12b}. \hs\ applied various filters to \citepossessive{henley12b} observations in order
to minimize the contamination from charge exchange (CX) emission from within the solar system, a
time-variable contaminant of SXRB spectra
\citep{cravens01,wargelin04,snowden04,koutroumpa07,fujimoto07,henley08a,ezoe10,carter11}.  In
addition, \hs\ excluded certain features from their sample (the Scorpius-Centaurus superbubble, the
Eridanus Enhancement, and the Magellanic Clouds). The \hs\ data set contains measurements for $\sim$4
times as many sight lines as \hskjm's data set, spanning the full range of Galactic longitudes.  Here
we give a brief overview of \hs's spectral modeling method and their halo results; see \hs\ for more
details, and for a comparison of their results with those from other recent studies of the SXRB.

\hs\ analyzed the SXRB spectrum for each sight line with a standard SXRB model, with components
representing the foreground, Galactic halo, and extragalactic background emission. The foreground
emission was constrained using shadowing data from the \rosat\ All-Sky Survey \citep{snowden00}.
For all but one sight line, \hs\ used a single-temperature ($1T$) collisional ionization equilibrium
(CIE) plasma model to model the halo emission, obtaining the X-ray temperature and emission measure
for the halo on each sight line. For the remaining sight line, \hs\ added another, hotter component
(with temperature $T \sim 11 \times 10^6~\K$), in order to model excess emission in the observed
spectrum around $\sim$1~\kev\ (see their Section~3.1.2).  For that sight line, we use the results
for the cooler component ($T \sim 2 \times 10^6~\K$).

\hs\ detected emission from $\sim$$(\mbox{2--3}) \times 10^6~\K$ halo plasma on 87 out of 110 sight
lines (79\%), with a median temperature of $2.2 \times 10^6~\K$,\footnote{For some sight lines,
  \hs\ were unable to constrain the halo temperature. In such cases, they fixed the temperature at
  $2.1 \times 10^6~\K$.} and a typical intrinsic 0.5--2.0~\kev\ surface brightness of
$(\mbox{1.1--2.3}) \times 10^{-12}~\flux\ \pdegsq$. On the remaining 23 sight lines, \hs\ give upper
limits for the halo surface brightness.

\citet{henley14d} compared a subset of \hs's results with a measurement of the Galactic halo
emission from an \xmm\ observation of a compact shadowing cloud, G225.60$-$66.40. The good agreement
between their measurement and that from the nearest \hs\ sight line led \citet{henley14d} to
conclude that \hs's measurements are not subject to systematic errors, and can confidently be used
to test models of the halo emission.

\section{GALACTIC FOUNTAIN MODELS}
\label{sec:FountainModel}

The \jm\ and \hill\ SN-driven ISM simulations were carried out using Flash,\footnote{Developed at
  the University of Chicago Center for Astrophysical Thermonuclear Flashes;
  http://flash.uchicago.edu/web/} a parallelized Eulerian hydrodynamical code with adaptive mesh
refinement (AMR). In each case, the model domain was a tall thin box extending to $z = \pm \zmax$,
with periodic boundary conditions on the vertical sides, and zero-gradient boundary conditions on
the upper and lower surfaces. The model domain was initialized with gas in hydrostatic
equilibrium. This gas was then heated and stirred stochastically by Type~Ia and core collapse SN
explosions, each with a frequency and $z$ distribution appropriate for the Milky Way in the vicinity
of the Sun. Each SN injected $10^{51}~\erg$ of energy into a small region of the grid. 60\%\ of the
core collapse SNe occurred in clusters of seven to 40 explosions, while the remaining SNe occurred
in isolation. The gas in the model domain was also subject to radiative cooling, and to diffuse
heating representing photoelectric heating of dust grains. The simulations were run at least long
enough to eradicate the initial conditions.  From our point of view, the most important feature of
these ISM models is that the SN heating drives a fountain of hot ($\ga$$10^6~\K$) X-ray-emissive gas
into the halo. For more details of the models, see \jm\ and \hill.

\hskjm\ tested the \jm\ ISM model, which was carried out in a $0.5 \times 0.5 \times 10~\kpc^3$
model domain, with $\zmax = 5~\kpc$. \hskjm\ found that the X-ray emission measures predicted by
this model were in good agreement with their observations, leading them to conclude that galactic
fountains are a major, possibly dominant, contributor to the hot X-ray emission in the \xmm\ band
(as noted in the Introduction). However, the \jm\ model overpredicted the observed X-ray
temperature.

It has subsequently been discovered that the boundary conditions at the upper and lower boundaries
of the \jm\ model domain led to an unphysical inflow of hot, high-pressure gas into the domain
(\citealt{maclow12,joung12a}; \hill). Early in the simulation, while the initial conditions were
still being eradicated, hot gas from SNe moved upward through and eventually off the domain, causing
the ghost cells just off the domain to be set to a high-temperature, high-pressure state. Subsequent
radiative cooling caused the halo pressure to drop, causing material to be drawn into the model
domain. The state of this inflowing material was determined by the state of the ghost cells, leading
to a hot, high-pressure inflow.  This inflow adversely affected the X-ray predictions derived from
the \jm\ model.

The newer \hill\ model used outflow-only boundary conditions, and was carried out in a much larger
domain---$1 \times 1 \times 40~\kpc^3$, with $\zmax = 20~\kpc$---and so does not suffer from the
unphysical inflow problem of the \jm\ model. There are a few additional differences from the
\jm\ model. First, the Type~Ia (core collapse) SN rate was slightly higher (lower) than that used in
\jm, though otherwise the SN heating was the same. Second, the grid initialization was slightly
different---\jm\ initialized their entire domain with gas at $10^4~\K$, whereas \hill\ initialized
their domain with gas at $1.15 \times 10^4$ and $1.15 \times 10^6$ below and above $|z| \approx
1~\kpc$, respectively (the pressure was continuous across the interface). Also, \hill\ employed a
higher gas surface mass density than \jm: 13.2 versus 7.5 \Msol\ $\pc^{-2}$. Finally, \hill\ ran
versions of their model that included a magnetic field---here, we examine versions with (model
\bx{50}) and without (model \bx0) a magnetic field. In model \bx{50}, the magnetic field was
initially horizontal and uniform in the $xy$ plane, with a magnitude of 6.5~\microgauss, but
decreased with height such that the ratio of magnetic and gas pressures was constant. Note that the
radiative cooling and diffuse heating were incorrectly applied in the original \hill\ simulations,
but corrected models were described in their erratum. Here, we use results from the corrected
simulations.

It should be noted that \hs\ found that the halo emission measure tends to increase toward the inner
Galaxy ($l=0\degr$). However, because the models' domains are tall thin boxes, we are unable to
determine how the model predictions would vary with Galactic longitude or latitude. Instead, we test
how well the models can reproduce the overall distributions of observed halo temperatures and
surface brightnesses.

\section{CHARACTERIZING THE FOUNTAIN MODEL X-RAY EMISSION}
\label{sec:FountainCharacterization}

\setlength{\tabcolsep}{3pt}
\tabletypesize{\scriptsize}
\begin{deluxetable*}{p{3mm}lrcr@{,}lcr@{,}lcr@{,}lcr@{,}lcr@{,}lcr@{,}l}
\tablecaption{Observed and Predicted Halo Temperatures and Surface Brightnesses, and Predicted Hot Gas Properties\label{tab:Quartiles}}
\tablehead{
 &                         &                  & \multicolumn{3}{c}{}            & \multicolumn{3}{c}{}                         & \multicolumn{12}{c}{Properties of hot ($T \ge 10^6~\K$) gas} \\
\cline{10-21}
 & \colhead{Obs. or Model} & \colhead{Time}   & \multicolumn{3}{c}{Temperature} & \multicolumn{3}{c}{\Stotal\tablenotemark{a}} & \multicolumn{3}{c}{\nmean}           & \multicolumn{3}{c}{\nrms}            & \multicolumn{3}{c}{$L$}   \\
 &                         & \colhead{(\Myr)} & \multicolumn{3}{c}{($10^6~\K$)} & \multicolumn{3}{c}{}                         & \multicolumn{3}{c}{($10^{-3}~\pcc$)} & \multicolumn{3}{c}{($10^{-3}~\pcc$)} & \multicolumn{3}{c}{(kpc)} \\
 & \colhead{(1)}           & \colhead{(2)}    & \multicolumn{3}{c}{(3)}         & \multicolumn{3}{c}{(4)}                      & \multicolumn{3}{c}{(5)}              & \multicolumn{3}{c}{(6)}              & \multicolumn{3}{c}{(7)}
}
\startdata
 1 & Observations         & \nodata    &  2.22 & (2.01  &  2.64)\tablenotemark{b} &  1.07 & (0.50  &  1.53)\tablenotemark{c} & \multicolumn{3}{c}{\nodata} & \multicolumn{3}{c}{\nodata} & \multicolumn{3}{c}{\nodata} \\
\hline
 2 & H12 \bx0     & 85         &  1.72 & (1.19  &  2.38)                  & 0.016 & (0.002 & 0.062)                  &  0.43 & (0.19  &  0.79) &  0.55 & (0.28  &  1.02) &  0.21 & (0.10  &  0.36) \\
 3 & H12 \bx0     & 135        &  1.67 & (1.10  &  2.34)                  & 0.006 & (0.000 & 0.031)                  &  0.35 & (0.15  &  0.66) &  0.45 & (0.17  &  0.85) &  0.16 & (0.04  &  0.34) \\
 4 & H12 \bx0     & 185        &  1.98 & (1.62  &  2.50)                  & 0.020 & (0.011 & 0.068)                  &  0.11 & (0.09  &  0.13) &  0.15 & (0.12  &  0.25) &  3.22 & (2.91  &  3.50) \\
 5 & H12 \bx0     & 235        &  1.74 & (1.33  &  2.73)                  & 0.012 & (0.003 & 0.035)                  & 0.042 & (0.036 & 0.050) & 0.078 & (0.046 & 0.123) &  7.98 & (7.05  &  8.94) \\
 6 & H12 \bx0     & 285        &  2.22 & (1.77  &  2.59)                  & 0.023 & (0.010 & 0.058)                  & 0.055 & (0.043 & 0.070) & 0.098 & (0.064 & 0.150) &  7.09 & (5.32  &  9.98) \\
 7 & H12 \bx0     & 335        &  1.88 & (1.68  &  2.17)                  & 0.010 & (0.006 & 0.040)                  & 0.037 & (0.035 & 0.044) & 0.059 & (0.040 & 0.112) &  9.76 & (8.29  & 11.17) \\
\hline
 8 & H12 \bx{50}  & 85         &  1.28 & (0.93  &  1.96)                  & 0.001 & (0.000 & 0.015)                  & 0.073 & (0.047 & 0.160) &  0.10 & (0.05  &  0.36) &  0.46 & (0.19  &  0.93) \\
 9 & H12 \bx{50}  & 135        &  1.35 & (1.01  &  2.03)                  & 0.004 & (0.001 & 0.035)                  &  0.13 & (0.10  &  0.20) &  0.17 & (0.11  &  0.40) &  0.63 & (0.43  &  0.89) \\
10 & H12 \bx{50}  & 185        &  1.52 & (1.02  &  2.17)                  & 0.005 & (0.001 & 0.029)                  & 0.098 & (0.040 & 0.140) &  0.15 & (0.09  &  0.29) &  1.11 & (0.70  &  1.59) \\
11 & H12 \bx{50}  & 235        &  1.86 & (1.35  &  2.78)                  & 0.015 & (0.002 & 0.048)                  & 0.037 & (0.025 & 0.056) &  0.11 & (0.04  &  0.18) &  4.14 & (3.64  &  4.88) \\
12 & H12 \bx{50}  & 285        &  1.92 & (1.61  &  2.79)                  & 0.012 & (0.003 & 0.041)                  & 0.021 & (0.018 & 0.027) & 0.066 & (0.041 & 0.109) &  9.36 & (9.07  &  9.66) \\
13 & H12 \bx{50}  & 335        &  1.93 & (1.60  &  2.63)                  & 0.011 & (0.002 & 0.046)                  & 0.017 & (0.014 & 0.022) & 0.060 & (0.026 & 0.111) & 11.06 & (9.66  & 12.03)
\enddata
\tablecomments{For each quantity, we have tabulated the median value, followed by the lower and upper quartiles in parentheses.
  For the model predictions, these quartiles were calculated from the sets of values obtained from the 242 model sight lines
  that we examined at each model epoch.
  Columns 3 and 4 contain halo temperatures and surface brightnesses, respectively.
  Columns~5--7 contain the mean electron densities, the \rms\ electron densities, and the path lengths of the hot gas
  along the model sight lines, respectively.}
\tablenotetext{a}{0.5--2.0~\kev\ surface brightness in $10^{-12}~\flux\ \pdegsq$.}
\tablenotetext{b}{Including only sight lines on which the temperature was free to vary.}
\tablenotetext{c}{Latitude-corrected value (assuming a plane parallel halo geometry), including non-detections at their best-fit values.}
\end{deluxetable*}

As in \hskjm, for a given model epoch we calculated halo X-ray spectra for 242 vertical sight lines,
looking upward and downward from the Galactic midplane. The vantage points for these sight lines
were arranged in an $11 \times 11$ grid in the Galactic midplane, with grid spacings of
$\approx$98~\pc. We used the \citet{raymond77} spectral code (updated by J.~C. Raymond \&
B.~W. Smith, 1993, private communication with R.~J. Edgar) to calculate the X-ray spectra, assuming
that the plasma is in CIE and is optically thin. We excluded material within 100~\pc\ of the
midplane from the emission calculations, as such material is not in the halo. Note that the SXRB
model used in the observational analysis (\hs) included a foreground component (in addition to the
halo component, the results for which we use here; Section~\ref{sec:Observations}). This foreground
component accounted for the observed emission from within $\sim$100~\pc\ of the midplane.

The true halo emission is likely from plasma with a range of temperatures; in the observational
analysis, this emission was characterized with a $1T$ plasma model (\hs). Similarly, the emission
predicted by the fountain models that we are examining here is from plasma with a range of
temperatures. Therefore, to ensure a like-with-like comparison of the models with the observations,
we characterized the predicted X-ray emission by creating synthetic \xmm\ observations of the SXRB,
and then analyzing the resulting spectra with the same SXRB model used in the observational
analysis. This method is described in full in \hskjm; here we give an overview, and point out the
differences from \hskjm.

For each model sight line, we combined the predicted halo X-ray emission with models for the
foreground emission, the extragalactic background, the instrumental fluorescence lines, and residual
soft proton contamination (\hs).  We folded the resulting spectrum through the \xmm\ response
function and added Poissonian noise corresponding to a typical field of view and exposure time from
\hs. The simulations also took into account the \xmm\ quiescent particle background.  We simulated a
MOS1 and a MOS2 spectrum for each model sight line, each of which we grouped such that there were at
least 25 counts per bin (as in \hs). We then fitted the grouped spectra with the input SXRB model,
but with the halo component replaced with a $1T$ CIE plasma model. We use the resulting best-fit
$1T$ halo model to calculate the intrinsic 0.5--2.0~\kev\ surface brightness, \Stotal. Thus,
for each model sight line, we obtained an X-ray temperature and surface brightness which
characterize the predicted X-ray spectrum, and which can be compared with the observed temperatures
and surface brightnesses. Note that the model surface brightnesses obtained in this way were
typically $\sim$20--40\%\ lower than those obtained directly from the model spectra. This is likely
because the $1T$ model used in the fitting cannot always accurately capture the entire model
spectrum, calculated from a multitemperature plasma.

We used the same foreground model as in \hskjm, but a different model for the extragalactic
background.  \hskjm\ used a single unbroken power-law \citep{chen97}, whereas we used the model from
\hs: a double broken power-law \citep{smith07a} rescaled to match the expected surface brightness of
the sources that fell below the source removal flux threshold \citep{moretti03,hickox06}. This
alteration in the extragalactic model resulted in a change in the typical level of soft proton
contamination; we adjusted our input model accordingly. The source removal flux threshold used in
\hs\ is lower than that used by \hskjm, resulting in more of the \xmm\ field of view being excluded
in the observational analysis. We therefore lowered the assumed field of view for the simulated
observations from 480 to 410~arcmin$^2$.  However, we kept the assumed exposure time at 15~\ks.

In order to ensure that the simulated spectra had adequate signal-to-noise to constrain the $1T$
halo model, we rescaled the input halo spectra such that they had a specified surface brightness.
We undid this rescaling at the end, by dividing the output emission measure by the same factor that
was used to multiply the input spectrum. In \hskjm, we rescaled the spectra to give a
0.4--2.0~\kev\ surface brightness of $2.06 \times 10^{-12}~\flux\ \pdegsq$. Here, we found that the
X-ray temperatures resulting from this procedure may depend weakly on the assumed surface brightness
used to rescale the spectra.  We therefore used three different values to rescale the spectra:
0.5--2.0~\kev\ surface brightnesses of $1.14 \times 10^{-12}$, $1.54 \times 10^{-12}$, and $2.34
\times 10^{-12}~\flux\ \pdegsq$ (these are the quartiles for sight lines on which
$\sim$$(\mbox{2--3})\times10^6~\K$ emission is detected; \hs, Table~2).

We subjected the halo and extragalactic components of the model to interstellar absorption.  The
assumed column density does not strongly affect the results, but here too we decided to use the
quartiles from \hs: $1.26 \times 10^{20}$, $1.63 \times 10^{20}$, and $2.12 \times 10^{20}~\pcmsq$
(cf. $1.7 \times 10^{20}~\pcmsq$ in \hskjm). Thus, each model sight line was characterized a total
of nine times.  For the comparison with the observations, we first combined the results obtained
with the different rescaling surface brightnesses and column densities.

\section{RESULTS}
\label{sec:Results}

\begin{figure*}
\centering
\plottwo{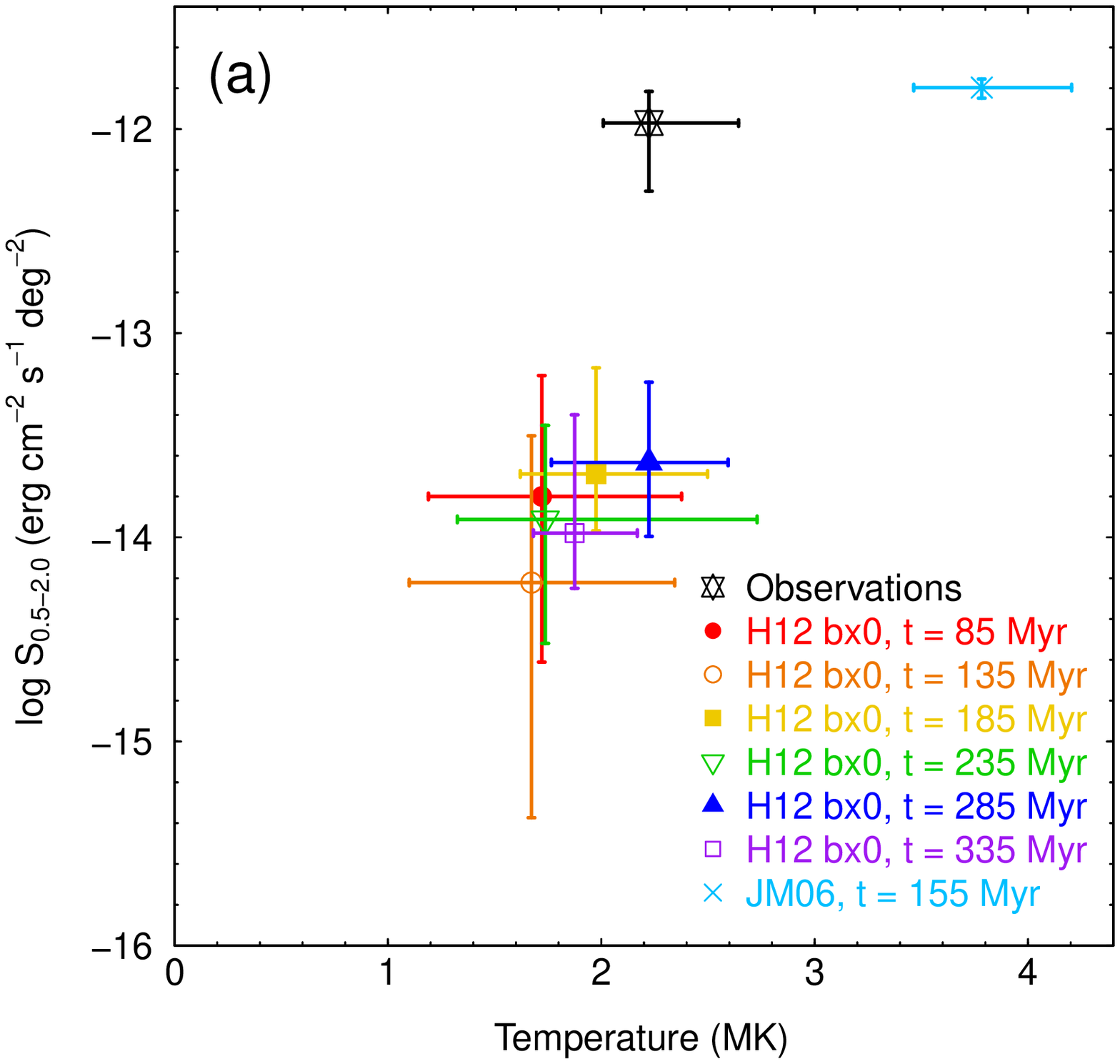}{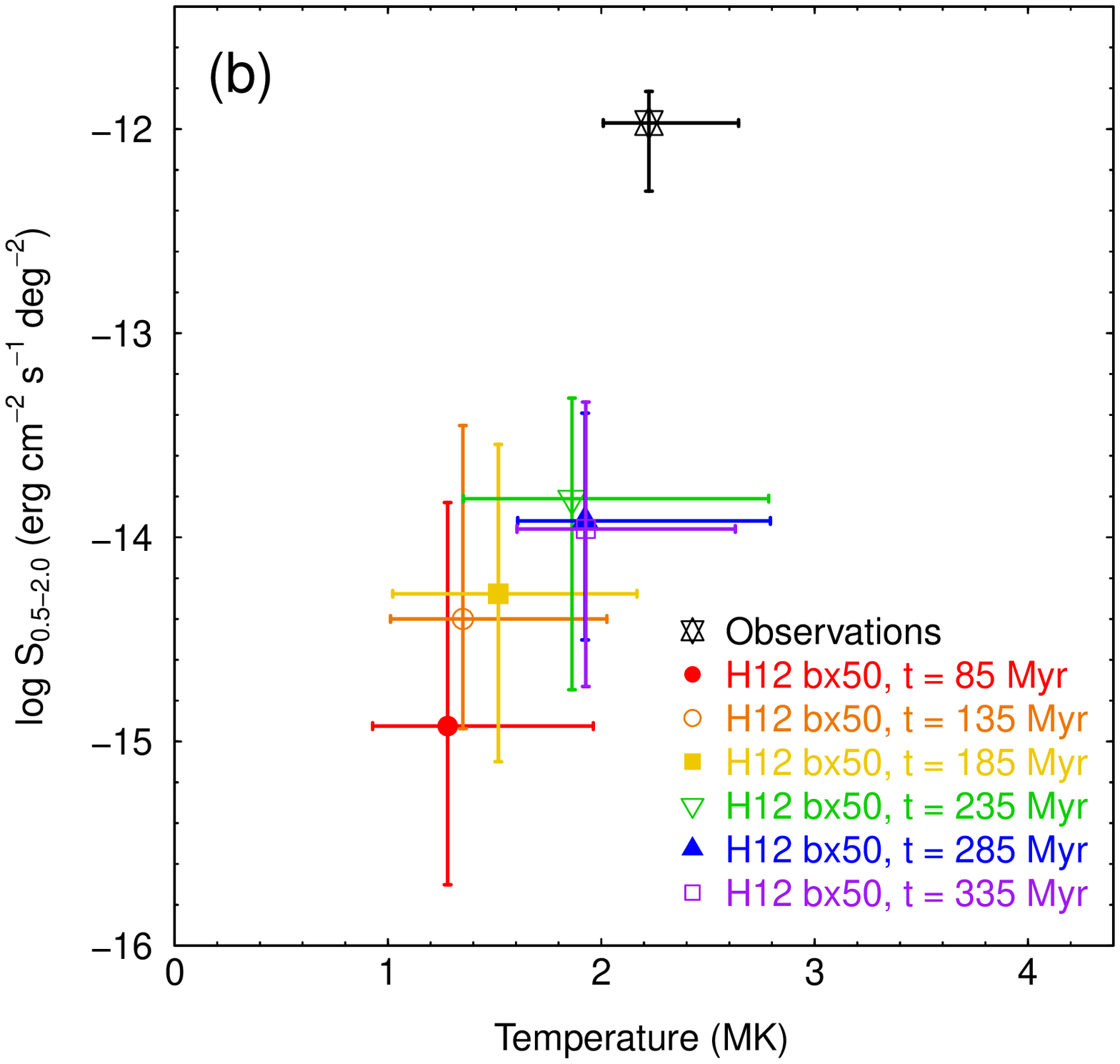}
\caption{Comparison of the \hill\ (a) \bx0 and (b) \bx{50} fountain model predictions with the
  \hs\ halo observations in the temperature--surface-brightness plane. The observations are plotted
  in black, while the other colors represent different epochs of the \hill\ models. For each plotted
  data point, the symbol indicates the medians, while the error bars indicate the lower and upper
  quartiles. For the observations, only sight lines on which the halo temperature was free to vary
  are included in the temperature data (see \hs). The observed surface brightnesses have been
  latitude-corrected assuming a plane parallel halo geometry, and non-detections are included at
  their best-fit values. In the \bx0 plot, we also show results for one epoch of the \jm\ model (the
  latest epoch examined by \hskjm).
  \label{fig:Obs+Model}}
\end{figure*}

\begin{figure*}
\centering
\plottwo{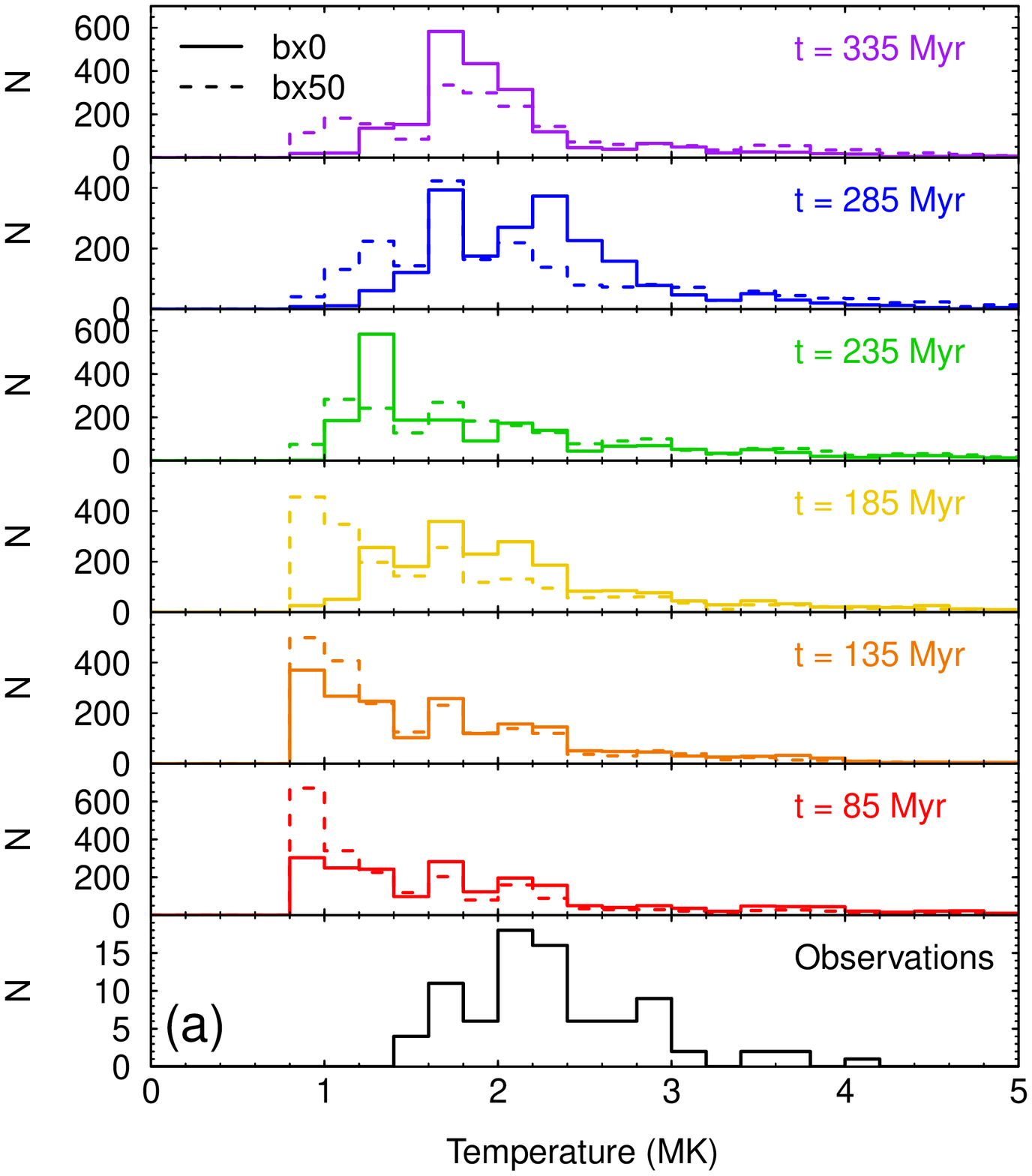}{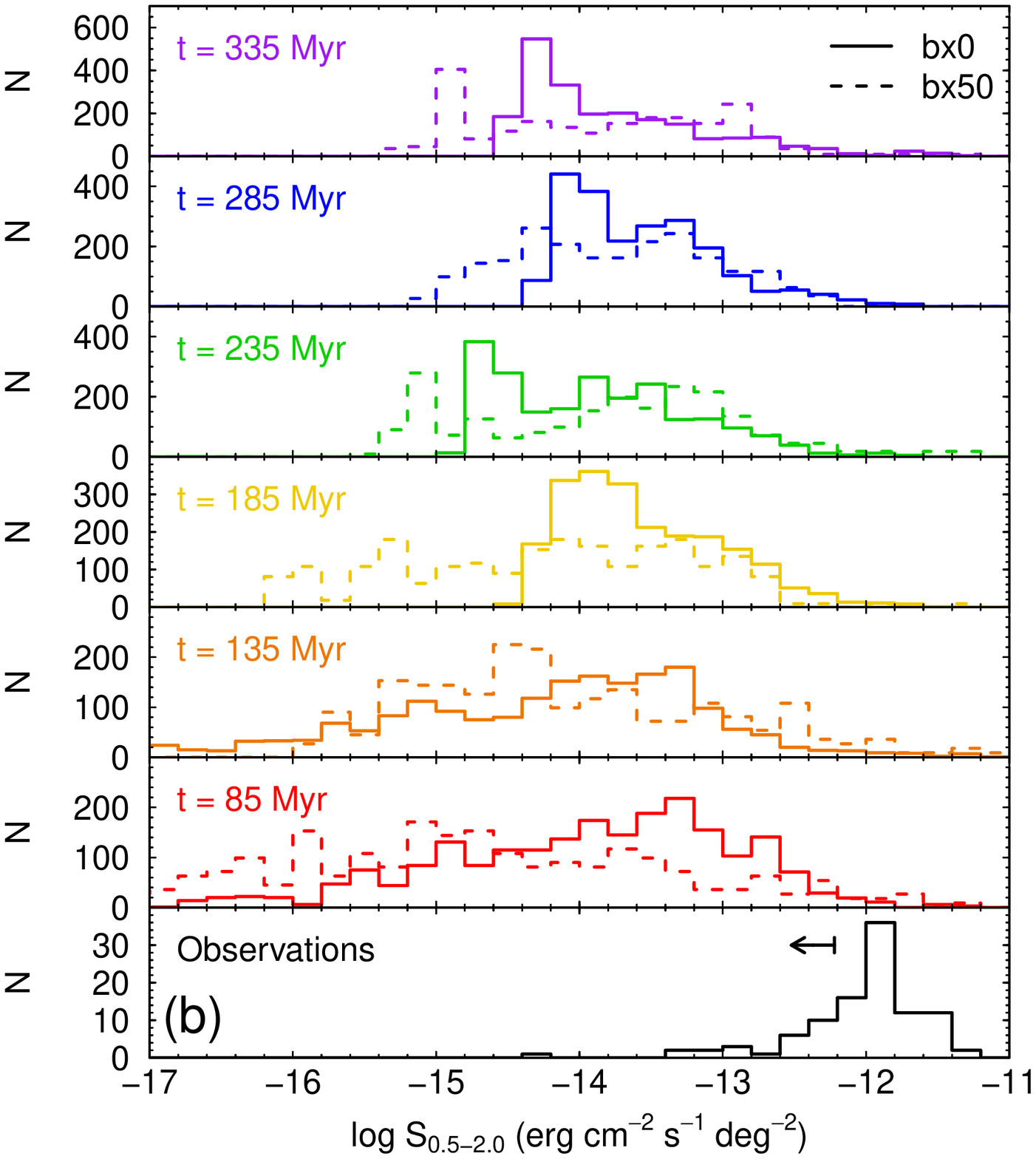}
\caption{Comparison of the observed and predicted distributions of halo (a) temperatures and (b)
  0.5--2.0~\kev\ surface brightnesses.
  In each plot, the bottom panel shows the observed distribution (\hs). Similarly to
  Figure~\ref{fig:Obs+Model}, only sight lines on which the halo temperature was free to vary are
  included in the observed temperature distribution, and the observed surface brightness
  distribution shows the latitude-corrected values with non-detections included at their best-fit
  values.
  The arrow in the bottom panel of plot (b) indicates the median upper limit on the latitude-corrected
  surface brightness from the sight lines on which halo emission was not detected.
  In the other panels, the solid and dashed histograms show the predicted distributions from
  different epochs of the \hill\ \bx0 and \bx{50} models, respectively. The colors representing the
  different model epochs match those used in Figure~\ref{fig:Obs+Model}.
  \label{fig:Obs+ModelDist}}
\end{figure*}

Figure~\ref{fig:Obs+Model} compares the X-ray predictions of the \hill\ fountain model with \hs's
halo observations in the temperature--surface-brightness plane. We show predictions from several
different epochs of the models (a) without and (b) with a magnetic field (models \bx0 and \bx{50},
respectively). For comparison with \hskjm, Figure~\ref{fig:Obs+Model}(a) also shows predictions from
a single epoch of the \jm\ model (the latest epoch examined by \hskjm).

Figure~\ref{fig:Obs+ModelDist} compares the predicted temperature and surface brightness
distributions of the \hill\ models with the observed distributions. Again, we show results from
several different epochs of models \bx0 and \bx{50}. The observed and predicted halo temperatures
and surface brightnesses are summarized in columns~3 and 4 of Table~\ref{tab:Quartiles},
respectively. The observations are in row~1, while the predictions from models \bx0 and \bx{50} are
in rows 2--7 and 8--13, respectively. Table~\ref{tab:Quartiles} also summarizes the properties of
the hot ($T \ge 10^6~\K$) gas from each epoch of the models---columns 5--7 contain the mean electron
densities, \nmean, the \rms\ electron densities, \nrms, and the path lengths, $L$, of this gas along
the model sight lines, respectively. Note that the \rms\ electron density is more useful than the
mean electron density for interpreting the X-ray emission predictions, since the emission measure
$\mathcal{E} = \nms L$.

The predictions from model \bx0 appear to undergo a slight oscillation in the
temperature--surface-brightness plane.  The median predicted X-ray temperature and surface
brightness oscillate with peak-to-peak amplitudes of $\sim$$0.3 \times 10^6~\K$ and $\sim$0.3~dex,
respectively, with a period of $\sim$100~\Myr. This oscillation may be related to the ``bouncing''
of the halo material reported by \hill.

At the earliest epochs of the \bx0 model shown here, the \rms\ density of the hot gas is relatively
high ($\sim$$5 \times 10^{-4}~\pcc$) and the path length through this gas is relatively short
($\sim$0.2~\kpc). At later epochs, the density is lower ($\sim$$1 \times 10^{-4}~\pcc$) but the path
length much longer (up to $\sim$10~\kpc). However, these changes are such that $\mathcal{E}$ remains
the same within a factor of $\sim$2 ($\sim$$5 \times 10^{-5}$ and $\sim$$1 \times
10^{-4}~\emismeas$, respectively, using the above-quoted densities and path lengths). (As an aside,
we note that the median \rms\ density and path length from the sight lines through the \jm\ model
are $9 \times 10^{-4}~\pcc$ and 4~\kpc, respectively, for the epoch plotted in
Figure~\ref{fig:Obs+Model}(a). Comparing these values with those from the later epochs of the
\hill\ \bx0 model implies that the main effect of the unphysical inflow in the \jm\ model is to
increase the density of the hot halo gas by an order of magnitude, and hence the X-ray surface
brightness by two orders of magnitude.)

The predictions from model \bx{50} do not oscillate, but instead there is a general increase in the
predicted X-ray temperatures and surface brightnesses from $t=85$ to $t=235~\Myr$---the medians
increase by $\sim$$0.6 \times 10^6~\K$ and an order of magnitude, respectively, over this time
period. The increase in brightness is mainly due to an order-of-magnitude increase in the path
length through the hot gas, due to a shock being driven upward through the halo.  The path length
through the hot gas continues to increase beyond $t=235~\Myr$, but more slowly than at earlier
epochs, as the shock slows down. The \rms\ density of the hot gas in the \bx{50} model decreases
from $t=135~\Myr$ onward, because the hot gas extends to greater heights, and thus includes
lower-density gas.  After $t=235~\Myr$, the model \bx{50} X-ray predictions are fairly steady,
although there is some variation in the shapes of the predicted distributions. At these later epochs
of the \bx{50} model, its predictions are similar to those from model \bx0.

Now that we have understood the \hill\ X-ray predictions in terms of the physical properties of the
hot gas in the model domains, we can compare these predictions with the \hs\ halo measurements. The
models generally underpredict the median observed halo temperature by $\sim$10--20\%, although the
predictions and observations agree within the observed sight line-to-sight line temperature
variation. However, the models significantly underpredict the halo surface brightness---the
difference is two orders of magnitude if we compare medians, and the predicted upper quartiles are
an order of magnitude less than the observed lower quartile. Note that there is more than enough
energy available in the model to power the observed X-ray emission in principle---the SN energy
injection rate in the \hill\ model is $1.1 \times 10^{39}~\ergps\ \kpc^{-2}$, whereas the observed
0.5--2.0~\kev\ luminosity is $\sim$$8 \times 10^{35}~\ergps\ \kpc^{-2}$ (\hs). However, in practice,
only $\sim$$10^{-5}$ of the energy from SNe in the model is radiated as 0.5--2.0~\kev\ photons from
the halo.

Despite the minor modifications to our method for obtaining the model predictions
(Section~\ref{sec:FountainCharacterization}), the results from the \jm\ model are consistent with
those in \hskjm---as in \hskjm, we find that this model matches the observed halo surface brightness
but overpredicts the halo temperature (Figure~\ref{fig:Obs+Model}(a)). However, as noted in
Section~\ref{sec:FountainModel}, the X-ray predictions from this model are unreliable, due to an
unphysical inflow of hot gas into the model domain.

\section{DISCUSSION}
\label{sec:Discussion}

Here we discuss possible reasons for the discrepancy between the \hill\ model predictions and \hs's
observations. First, we consider the impact of variations in the SN rate
(Section~\ref{subsec:SNRate}).  We then consider the possibility that we are underestimating the
emission from the halo material in the \hill\ model, either because we assume that the halo gas is
in CIE (Section~\ref{subsec:NEI}), or because we are underestimating the emission from interfaces
between hot and cold gas, due to thermal conduction not being included in the hydrodynamical model
(Section~\ref{subsubsec:Conduction}) and CX not being included in the emission model
(Section~\ref{subsubsec:CX}). We then consider the possibility that cosmic rays (CRs) play a role in
driving material out of the disk, meaning that the \hill\ model may underestimate the amount of hot
material in the halo (Section~\ref{subsec:CRDriving}).  Finally, we consider the role that a more
extended halo of hot gas, predicted by galaxy formation models, may play in producing the observed
X-ray emission (Section~\ref{subsec:Extended}).

\subsection{Supernova Rate}
\label{subsec:SNRate}

We first consider the impact of our chosen model parameters, such as the SN rate and the gas surface
mass density. \Citet{avillez04} and \citet{joung09} have each explored variations in the SN rate in
similar models; \citeauthor{joung09}\ correspondingly varied the gas surface mass density following
the Kennicutt-Schmidt law \citep{kennicutt98}. \Citet{avillez04} found that the hot gas filling
fraction increases somewhat with SN rate; \citet{joung09} found that the turbulent pressure and
thermal pressure track each other. Both found that the temperature of the hot gas increases somewhat
in higher SN rate models with relatively little change in the density of the hot gas (see Figures~2
and 3 of \citealt{avillez04} and Figure~2 of \citealt{joung09}).  We thus suspect that an increased
SN rate would increase the X-ray temperature and, as a result, the X-ray surface brightness. A
higher hot gas filling fraction and density would also increase the surface brightness, by
increasing the emission measure. Because the \citet{joung09} models are not directly comparable to
the H12 models (see Section~\ref{sec:FountainModel}), a quantitative estimate of this effect would
require running versions of the H12 models with varied SN rates. This is beyond the scope of this
paper.

Local variations in the SN history due to the pseudorandom SN distribution may also impact the
observed properties. The consideration of multiple time steps in a single model addresses this
source of uncertainty to some extent. However, the variations in both temperature and emission
measure are relatively small over the course of the runs (Figures~\ref{fig:Obs+Model} and
\ref{fig:Obs+ModelDist}).

\subsection{Non-equilibrium Ionization}
\label{subsec:NEI}

We now consider the possibility that the \hill\ model underpredicts the observed halo X-ray emission
because we assumed that the X-ray-emitting plasma was in CIE when calculating the X-ray spectral
predictions (Section~\ref{sec:FountainCharacterization}). In reality, the plasma in the halo may be
overionized (i.e., the ionization temperature exceeds the kinetic temperature) as a result of
radiative or adiabatic cooling. This would result in recombination emission from cool gas
which is essentially non-emissive if we assume CIE \citep{breitschwerdt94,avillez12a}.  Hence, by
assuming CIE, we may be underestimating the X-ray emission from the halo plasma in the
\hill\ model. In addition, non-equilibrium ionization (NEI) would affect the radiative cooling rate,
which would in turn affect the temperature structure of the halo in the hydrodynamical models---this
too could affect the X-ray predictions.

It is not possible to calculate the degree of overionization, and hence the amount of recombination
emission to include in the X-ray spectral predictions, when post-processing the
\hill\ hydrodynamical data, as Lagrangian temperature histories are not available. Instead, one
needs to trace self-consistently the ionization evolution of the relevant elements during the course
of the hydrodynamical simulation. While such simulations do exist \citep{avillez12a,avillez12b},
X-ray spectral predictions that can be compared directly with \hs's observations are not currently
available. Note that, although an overionized recombining plasma produces a very different emission
spectrum from the CIE plasma models used in \hs's \xmm\ analysis (free-bound versus line emission),
future predictions from NEI ISM models could still be compared with \hs's observational results, if
such predictions are first characterized using the method described in
Section~\ref{sec:FountainCharacterization}.

Although detailed X-ray spectral predictions for a recombining halo plasma are not currently
available, we can estimate by how much taking into account NEI would increase the predicted X-ray
surface brightness of the \hill\ model. For this calculation, we used the code described in
\citet{shelton98} to follow the ionization evolution of a stationary parcel of plasma initially in
CIE cooling isobarically from $3 \times 10^6~\K$.\footnote{While similar calculations have been
  carried out previously \citep[e.g.,][]{shapiro76b,avillez12a}, the results are not presented in a
  form that can easily be applied to the 0.5--2.0~\kev\ \xmm\ band.} At each step in the
calculation, the code takes into account the non-equilibrium ion populations in the plasma when
calculating the radiative cooling function and the emergent X-ray spectrum. We find that, when the
plasma has cooled to $3 \times 10^5~\K$, the 0.5--2.0~\kev\ emission is $\sim$3,000 times as bright
as that from a CIE plasma at the same temperature. However, this overionized, recombining plasma is
$\sim$17,000 times fainter than the original $T = 3 \times 10^6~\K$ CIE plasma. In the \hill\ model,
the emission measure of gas with $T = (\mbox{2--4}) \times 10^5~\K$ is typically similar to (within
a factor of $\sim$5) the emission measure of gas with $T \ge 1 \times 10^6~\K$.\footnote{Note that
  the flatness of the mass-weighted temperature distributions below $\logT \sim 6$ indicates that
  there are similar quantities of $\sim$$3 \times 10^5$ and $\sim$$1 \times 10^6~\K$ gas in the
  \hill\ model domains (see Figure~5 of the \hill\ erratum).} This calculation therefore implies
that the overionized cooled halo plasma would be much fainter than the hot ($T \ge 10^6~\K$) halo
plasma, and so taking into account overionization in the cooled halo plasma would not significantly
increase the total X-ray surface brightness predicted by the \hill\ model.

\subsection{Emission from Interfaces}
\label{subsec:Interfaces}

\subsubsection{Effect of Thermal Conduction}
\label{subsubsec:Conduction}

We now explore the possibility that, because the \hill\ model does not include thermal conduction,
we are underestimating the contribution to the emission from interfaces between tenuous hot gas ($T
\ga 10^6~\K$) and denser, cooler gas ($T \la 10^4~\K$). Within such interfaces there exists
X-ray-emissive gas that is denser, and as a result brighter per unit volume, than the diffuse hot
gas. These interfaces are typically $\sim$10--70~\pc\ thick along the line of sight in the present
model (note that these interfaces are not well resolved in the halo, where the resolution is
typically 16 or 32~\pc\ in the hot gas). Thermal conduction would tend to broaden these interfaces
until their widths are approximately equal to the Field length \citep{begelman90b},
\begin{equation}
  \lambda_\mathrm{F} = \left( \frac{\kappa T}{n^2 \Lambda} \right)^{1/2},
  \label{eq:FieldLength}
\end{equation}
where $n$ is the number density, $\kappa = 5.6 \times 10^{-7}
(T/\K)^{5/2}\ \erg\ \ps\ \K^{-1}\ \cm^{-1}$ is the thermal conductivity \citep{draine84}, and
$\Lambda$ is the radiative cooling function \citep[and updates]{raymond77}.\footnote{In the
  definition of $\lambda_\mathrm{F}$ in \citet{begelman90b}, the second term in the denominator of
  Equation~(\ref{eq:FieldLength}) is $\mathcal{L}_M \equiv \max(\Lambda, \Gamma/n)$, where $\Gamma$ is
  the diffuse heating rate. However, at the temperatures in the interfaces considered here, $\Gamma
  = 0$ (\hill), and so $\mathcal{L}_M = \Lambda$.}  Taking values from the midpoints of the
interfaces in the \hill\ model (where the temperatures and densities are typically $\sim$$9 \times
10^4$--$7 \times 10^5~\K$ and $\sim$$2 \times 10^{-4}$--$2 \times 10^{-2}~\pcc$, respectively), we
obtain Field lengths typically in the range $\sim$1--600~\pc\ (though for some interfaces,
$\lambda_\mathrm{F} < 0.1~\pc$ or $>$1~\kpc). For approximately half of the interfaces in the
\hill\ model, $\lambda_\mathrm{F}$ exceeds the interface width, implying that thermal conduction
would tend to broaden these interfaces. All other things being equal, increasing the width of such
an interface increases the path length through the denser X-ray-emissive gas, and so including
thermal conduction would be expected to boost the emission from interfaces.

\begin{figure}
\plotone{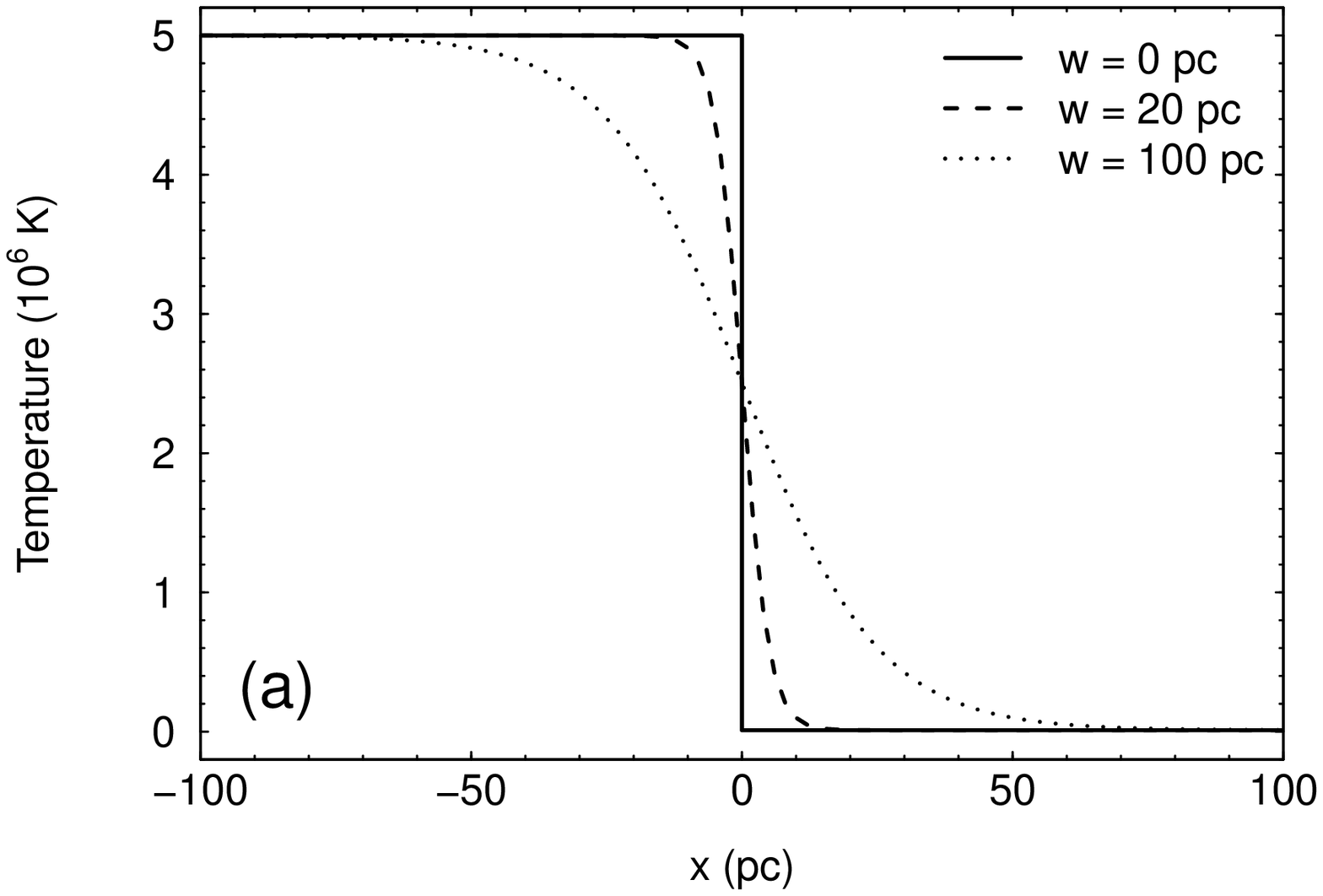}
\plotone{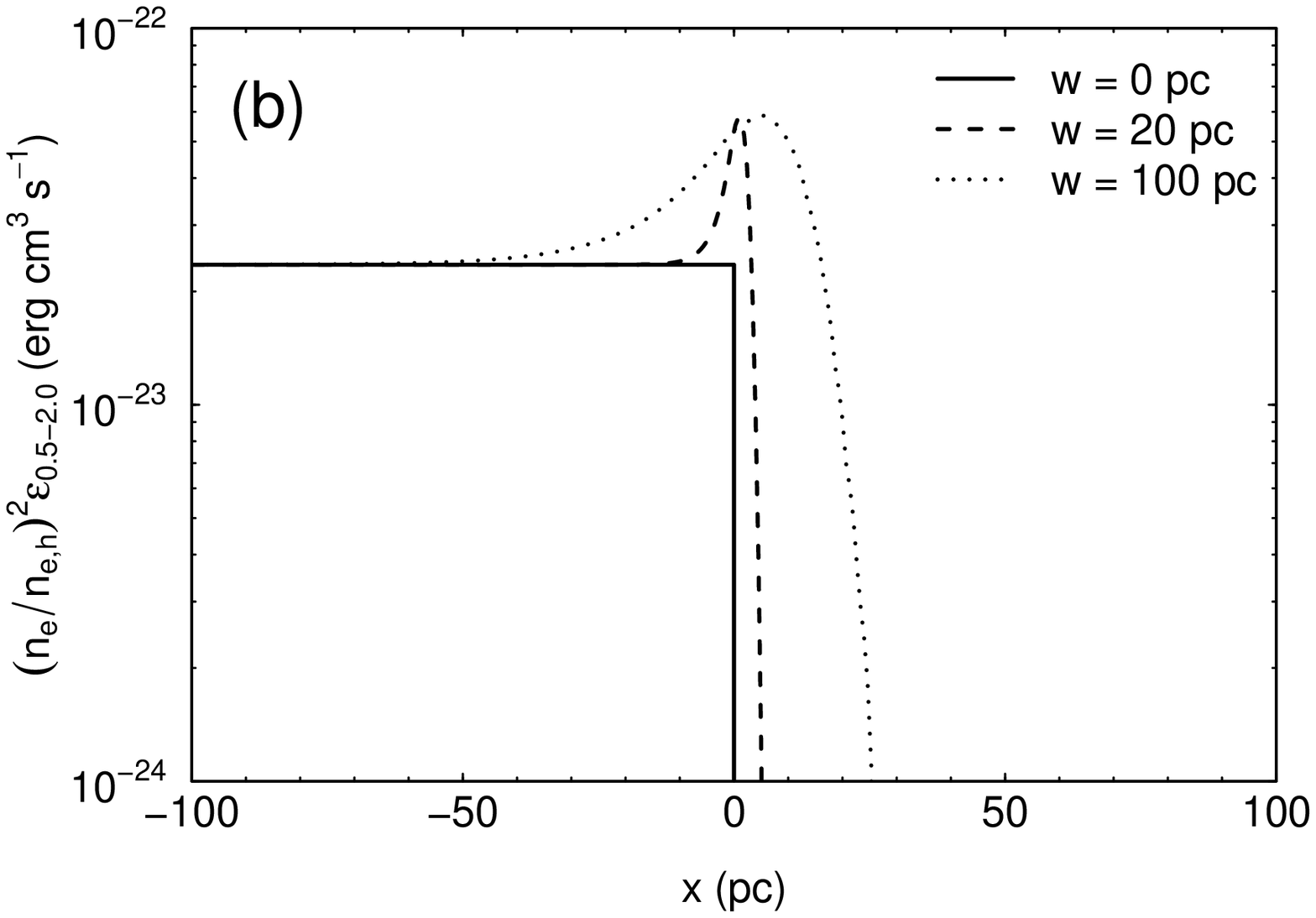}
\plotone{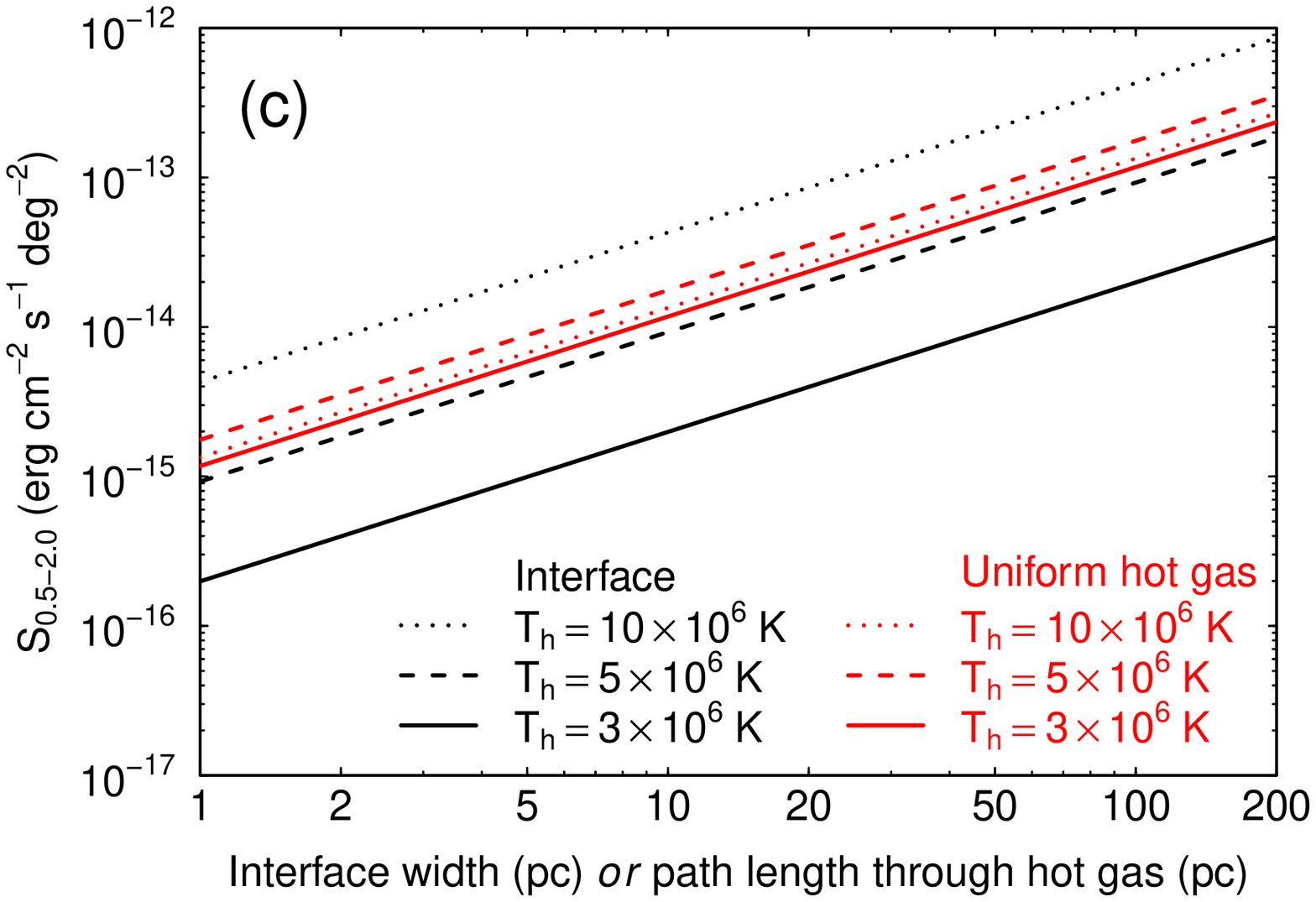}
\caption{(a) Temperature profiles for the interfaces between hot and cold gas described by
  Equation~(\ref{eq:InterfaceT}), with $\Th = 5 \times 10^6~\K$ and $\Tc = 1 \times 10^4~\K$, for
  three different interface widths (\textit{solid}: $w=0~\pc$; \textit{dashed}: $w=20~\pc$;
  \textit{dotted}: $w=100~\pc$).
  (b) Profiles of the 0.5--2.0~\kev\ emission for these model interfaces, where the plasma
  emissivity, $\varepsilon_{0.5-2.0}$, is a function of the temperature given by
  Equation~(\ref{eq:InterfaceT}), and the electron density, \Ne, is given by
  Equation~(\ref{eq:Interfacen}). The plot shows profiles for the same values of \Th\ and
  \Tc\ and for the same interface widths as in panel (a).
  (c): 0.5--2.0~\kev\ surface brightnesses of these model interfaces
  (Equation~(\ref{eq:InterfaceSB}); \textit{black curves}) and of regions of uniform hot gas of
  temperature \Th\ (\textit{red curves}), as functions of interface width, $w$, and of path length
  through the hot gas, respectively. In each case, results are shown for $\Th = 3 \times 10^6$
  (\textit{solid}), $5 \times 10^6$ (\textit{dashed}), and $10 \times 10^6~\K$ (\textit{dotted}),
  assuming $\neh = 10^{-3}~\pcc$.
  \label{fig:Interface}}
\end{figure}

We investigated by how much the broadening of interfaces by thermal conduction could increase the
X-ray emission by considering smoothly varying model interfaces of width $w$ between hot ($\Th =
\mbox{several} \times 10^6~\K$) and cold ($\Tc = 10^4~\K$) gas in pressure balance. In our interface
model, the temperature across the interface varies with position $x$ as
\begin{equation}
  T(x,w) = \frac{\Tc + \Th}{2} + \frac{\Tc - \Th}{2} \tanh \left( \frac{4x}{w} \right),
  \label{eq:InterfaceT}
\end{equation}
where the interface center is located at $x=0$. As the interface is in pressure balance, the
electron density is
\begin{equation}
  \Ne(x,w) = \neh \frac{\Th}{T(x,w)},
  \label{eq:Interfacen}
\end{equation}
where \neh\ is the density in the hot gas. Temperature profiles for three example values of $w$ are
shown in Figure~\ref{fig:Interface}(a). Corresponding profiles of the 0.5--2.0~\kev\ X-ray emission
(normalized to \neh), $(\Ne/\neh)^2 \varepsilon_{0.5-2.0}(T)$, where $\varepsilon_{0.5-2.0}(T)$ is
the plasma emissivity, are shown in Figure~\ref{fig:Interface}(b). As can be seen, an interface with
$w > 0$ is locally brighter than a zero-width interface between $x \approx -w/2$ and $x \approx
+w/4$.

By integrating emission profiles like those shown in Figure~\ref{fig:Interface}(b) with respect to
$x$, we can calculate the X-ray surface brightness of an interface described by
Equations~(\ref{eq:InterfaceT}) and (\ref{eq:Interfacen}) as a function of interface width $w$,
\begin{align}
  \Stotal(w) = \frac{1}{4\pi} \int^\infty_{-\infty} \Big\{ & \Ne^2(x,w) \varepsilon_{0.5-2.0}(T[x,w]) - \nonumber \\
                                                           & \Ne^2(x,0) \varepsilon_{0.5-2.0}(T[x,0]) \Big\} dx
  \label{eq:InterfaceSB}
\end{align}
(for a sight line looking perpendicular to the interface). Note that, in the above expression, we
subtract off the emission from a zero-width interface, so $\Stotal(w)$ is the increase in surface
brightness due to increasing an interface's width from zero to $w$.  Note also that, since the
integrand in Equation~(\ref{eq:InterfaceSB}) is a function of $x/w$ (see
Equations~(\ref{eq:InterfaceT}) and (\ref{eq:Interfacen})), $\Stotal \propto w$. The surface
brightnesses obtained from Equation~(\ref{eq:InterfaceSB}) are shown by the black curves in
Figure~\ref{fig:Interface}(c), for three different values of \Th.

While increasing the widths of the interfaces would increase their surface brightness, in practice
the increase in brightness cannot account for the discrepancy between the \hill\ predictions and the
\hs\ observations. This is because the emission from the diffuse hot gas tends to dominate over that
from the interfaces between the hot gas and cooler gas, as we now demonstrate.  The red curves in
Figure~\ref{fig:Interface} shows the surface brightnesses of uniform hot gas as functions of path
length through the hot gas, for the same three values of \Th\ used for the black curves.  For $\Th
\la 5 \times 10^6~\K$, the surface brightness of an interface of a given width is less than the
surface brightness of uniform hot gas of the same extent.  As the regions of diffuse hot gas will
likely be larger than the interfaces at their edges, the emission from the diffuse hot gas will tend
to dominate. For example, consider a region of diffuse $5 \times 10^6~\K$ gas 500~\pc\ in extent
with zero-width interfaces at its edges. If thermal conduction were to increase the widths of those
interfaces from zero to 100~\pc, the total X-ray surface brightness would increase by only
$\sim$20\%. Therefore, increasing the widths of the interfaces in the model (by including thermal
conduction) would not counteract the two-order-of-magnitude difference in brightness between the
\hill\ predictions and the \hs\ \xmm\ observations.

\subsubsection{Charge Exchange}
\label{subsubsec:CX}

The above discussion considered only emission resulting from collisional excitation of the gas in an
interface. However, CX reactions between ions from the hot side of an interface and neutral H and He
atoms from the cold side could also contribute to the emission. To estimate the importance of CX
emission, we used Equation~(1) from \citet{lallement04b}. This gives the path length, \LCX, through
a hot gas for which the thermal emission from the hot gas is equal in brightness to the CX emission
from the two interfaces at either end of the hot gas. Assuming the interfaces are observed at normal
incidence, this path length is
\begin{equation}
  \LCX = 0.06 \frac{\epsilon \alpha \nc V_{100}}{\chi \Ne^2}~\pc,
  \label{eq:LCX}
\end{equation}
where $\epsilon$ is the ratio between the CX probability and the collisional ionization probability in
the hot gas, $\alpha$ is the ratio of the global emissivity CX cross-section, $\Sigma$, to that
assumed by \citet{lallement04b} ($\Sigma = 6 \times 10^{-19}~\kev\ \cm^2$, appropriate for solar
wind CX emission in the 0.1--0.5~\kev\ band), $\chi$ is the ratio of the hot gas emissivity to that
assumed by \citeauthor{lallement04b} ($5.8 \times 10^{-14}~\kev\ \cm^3\ \ps$), \nc\ and \Ne\ are the
number densities of the cold and hot gas, respectively, and $V_{100}$ is the relative speed of the
ions and the neutrals in units of 100~\kmps.

To estimate \LCX, we assumed that the hot gas has a temperature of $3 \times 10^6~\K$, and used
$\epsilon = 0.2$ (from the $V=100~\kmps$ curve in Figure~1 of \citealt{lallement04b}) and $\chi =
0.17$ (the 0.5--2.0~\kev\ emissivity of a $3 \times 10^6~\K$ plasma is $9.8 \times
10^{-15}~\kev\ \cm^3\ \ps$; \citealt{raymond77} and updates). In the absence of suitable CX emission
data for the \xmm\ band, we assumed $\alpha \sim 1$.\footnote{Although we are considering a
  higher-energy band than \citet{lallement04b} (0.5--2.0 versus 0.1--0.5~\kev), and hence CX
  emission from a different set of lines, we are assuming here that the sum of the abundances of the
  relevant ions and the typical CX cross-section and line yield are similar to the values that
  yielded \citeauthor{lallement04b}'s assumed value of $\Sigma$. The line energies will of course be
  higher in the band that we are considering, which would tend to increase $\Sigma$.  However, as
  the CX emission in the \xmm\ band is likely dominated by oxygen \Kalpha\ emission near 0.6~\kev,
  the typical line energies in the two bands will be within a factor of a few of each other, and so
  $\alpha \sim 1$ should be a reasonable assumption.} We used a typical hot gas density of $5 \times
10^{-5}~\pcc$ (Table~\ref{tab:Quartiles}, column~5), and we assumed that the hot gas is in pressure
equilibrium with cold gas with temperature $3 \times 10^4~\K$ (i.e., $\nc = 100\Ne$). Finally, we
assumed that the relative motion of the ions and neutrals is dominated by thermal motion, and so
used $V_{100} \sim 1$ (cf.\ the mean speed of oxygen ions in a $3 \times 10^6~\K$ plasma is
60~\kmps).

Using the above values in Equation~(\ref{eq:LCX}), we find $\LCX \sim 140~\kpc$, i.e., a region of
hot ($T=3\times10^6~\K$, $\Ne=5\times10^{-5}~\pcc$) gas in pressure equilibrium with cooler
($3\times10^4~\K$) gas would have to be $\sim$140~\kpc\ in extent in order for its
0.5--2.0~\kev\ thermal emission to be as bright as the CX emission from the interfaces bounding the
gas. In contrast, the path lengths through the hot gas in the \hill\ model are typically a few
kiloparsecs (Table~\ref{tab:Quartiles}, column~7), implying that CX emission may be up to two orders
of magnitude brighter than the thermal emission from the hot gas. It is therefore possible that CX
emission could account for much of the shortfall between the current predictions from the
\hill\ model and the observed halo surface brightness. However, from this simple estimate we cannot
definitively conclude that most of the observed halo emission is due to CX---more detailed spectral
calculations are needed to determine how much CX emission the \hill\ model produces. These
calculations would have to be carried out for each hot-cold interface in the model individually,
taking into account the temperature of the hot gas (which affects the populations of the ions
undergoing CX reactions) and the densities of the hot and cold gas (which affect the overall
brightness of the CX emission), and using CX cross-section and line yield data suitable for emission
in the \xmm\ band. Such calculations are beyond the scope of this paper.

If CX emission is indeed a major contributor to the observed halo X-ray emission, this would mean
that the emission measure of the hot halo gas is smaller than previously thought (e.g.,
$\sim$$\mbox{(0.4--7)} \times 10^{-3}~\emismeas$; \hs). This would have important implications for
the results of joint emission-absorption analyses of the halo, in which emission measurements are
combined with ion column density measurements to infer the density and extent of the halo. In such
an analysis, the extent of the hot halo scales as $N^2/\mathcal{E}$, where $N$ and $\mathcal{E}$ are
the column density and emission measure of the hot gas, respectively. The density scales as
$\mathcal{E}/N$, and so for a spherical halo, the gas mass scales as $\mathcal{E}/N \times
(N^2/\mathcal{E})^3 = N^5/\mathcal{E}^2$. Therefore, if the presence of CX means that the halo
emission measure is overestimated, the extent and mass of the hot halo inferred from joint
emission-absorption analyses will be underestimated.

\subsection{Cosmic Ray Driving}
\label{subsec:CRDriving}

In the \hill\ model, material is driven from the disk into the halo solely by the thermal pressure
of SN-heated gas. However, CRs may also play an important role in driving outflows
from galactic disks \citep{breitschwerdt91}. \citet{everett08} showed that a CR-driven galactic wind
(modeled in one dimension) provided a better fit to the diffuse 3/4~\kev\ emission observed toward
the inner Galaxy ($-30\degr<l<30\degr$) than a static polytropic model. \citet{salem14}, meanwhile,
used three-dimensional AMR simulations to study CR-driven outflows. They showed that CR driving led
to significant baryonic mass loss from the disk of their model galaxy, in contrast to a model
without CR driving, in which there was no such mass loss. In addition, \citet{salem14} showed that
including CR diffusion (as opposed to just having the CRs advect along with the gas flow) was
important for driving the outflow from the disk.

In the context of the present study, including CR driving would be expected to result in more
material being transported from the disk into the halo than in the \hill\ model, thus potentially
increasing the halo's X-ray surface brightness. However, the X-ray emission from such a CR-driven
outflow also depends on its temperature structure. \citet{booth13} found that CR driving results in
cooler outflows than pure thermal-pressure driving, but they chose a feedback implementation
equivalent to the ``energy only'' runs in \citet{agertz13}, which minimizes or eliminates hot gas
production by SN explosions in dense gas (see Figure~6 in \citeauthor{agertz13}). As a result, their
prediction is only a lower limit on the true temperature. A model similar to the \hill\ model that
incorporates CRs is currently under development (P.\ Girichidis et al.\ 2014, in preparation).  The
X-ray predictions from this new model will help determine the role of CR driving in supplying the
hot halo gas observed in emission.

\subsection{Role of an Extended Galactic Halo}
\label{subsec:Extended}

Finally, we consider how an extended halo of hot gas ($\ga$100~\kpc\ in extent) might affect the
\hill\ model predictions.  The emission from the \hill\ model comes mostly from within a few
kiloparsecs of the Galactic midplane. This is in part due to the low densities far above the disk in
the model (typically $\la$$3 \times 10^{-5}~\pcc$ above 10~\kpc). In contrast, there is indirect
evidence (from the lack of gas in satellite galaxies and the confinement of high-velocity clouds)
for higher density halo gas far from the disk \citep[and references therein]{fang13}. For example, a
model of an extended non-isothermal halo in hydrostatic equilibrium with the Galaxy's dark matter
\citep{maller04}, which is consistent with the observed X-ray emission and pulsar dispersion measure
data, and with the aforementioned indirect evidence \citep{fang13}, has a density exceeding
$10^{-4}~\pcc$ out to $\sim$100~\kpc\ (MB model in Figure~1 of \citealt{fang13}). Such extended hot
halos are also predicted by disk galaxy formation models \citep[e.g.,][]{crain10}.

If the Milky Way's extended halo consists of low-metallicity material accreted from the
intergalactic medium, then in itself it would not be X-ray bright. However, from smoothed particle
hydrodynamics (SPH) simulations of galaxy formation, \citet{crain13} found that the X-ray emission
from their model galactic halos was produced by metals that were transported out of the ISM being
collisionally excited by hot electrons in low-metallicity accreted gas. Hence, a hot,
low-metallicity halo of accreted material could boost the X-ray emission from the fountains in the
\hill\ model, by increasing the population of electrons available to excite the ions in the
fountains.  In addition, if there have been previous episodes of starburst activity in the Milky
Way, these could have enriched the extended halo with metals, potentially making the extended halo
intrinsically X-ray emissive. The arbitrary inflow at the boundaries of the \jm\ model in fact
raised the high-altitude densities above $10^{-4}~\pcc$. As found in \hskjm, this did indeed lead to
X-ray surface brightnesses comparable to the observed values. If the extended halo is indeed
intrinsically X-ray emissive, its emission would have to be added to that predicted by the
\hill\ fountain model to get the total predicted halo emission. Note, however, that the observed
halo emission is patchy, exhibiting large sight line-to-sight line variation (\citealt{yoshino09};
\hs). An extended halo model may have difficulty explaining this patchiness.

Predictions from hydrodynamical models of galaxy formation are needed to test the role played
by an extended halo in producing the X-ray emission observed from the Milky Way's halo. We plan to
examine such predictions in a subsequent paper.

\section{SUMMARY AND CONCLUSIONS}
\label{sec:Summary}

We have compared the X-ray emission predictions of a magnetohydrodynamical model of the SN-driven
ISM (\hill) with \xmm\ measurements of the Galactic halo's emission (\hs). This model significantly
underpredicts the halo's X-ray surface brightness (by two orders of magnitude, when we compare the
medians of the predicted and observed values; Section~\ref{sec:Results}). Including an interstellar
magnetic field does not significantly affect these X-ray predictions.

We explored possible reasons for the discrepancy between the \hill\ model predictions and \hs's
\xmm\ observations. Assuming CIE may in principle underestimate the emission from the \hill\ model
halo, but in practice this is unlikely to have a significant effect (Section~\ref{subsec:NEI}). We
also found that the discrepancy could not be explained by the emission from interfaces in the
\hill\ model being underestimated due to a lack of thermal conduction in the model
(Section~\ref{subsubsec:Conduction}). However, CX emission from such interfaces (not included in the
present emission model) could greatly increase the predicted X-ray surface brightness, though
detailed spectral calculations are needed to confirm this (Section~\ref{subsubsec:CX}).  (If CX
emission is a major contributor to the observed halo emission, then the hot gas emission measure is
less than previously thought, with the consequence that the path length and mass of the hot gas
calculated from algebraic combinations of the emission measure and ion column density would be
revised upwards.)  In addition, CR driving of a wind could increase the amount of X-ray-emissive
material in the halo (Section~\ref{subsec:CRDriving}), and an extended hot halo of accreted
material, while not intrinsically X-ray bright, may supply hot electrons that could increase the
predicted X-ray emission from galactic fountains (Section~\ref{subsec:Extended}).

In conclusion, the faintness of the \hill\ model relative to the observed surface brightness implies
that thermal emission from classical galactic fountains is not a major source of the halo's X-ray
emission. This is in contrast to the conclusion of \hskjm, which was based on the \jm\ model (the
X-ray predictions from which are now known to be incorrect). Our results indicate that additional
physical processes need to be included in halo models.  Two plausible possibilities are the effects
of CR driving on the fountain, and extended hot halos from the galaxy formation process. In
addition, CX may be an important contributor to the observed emission.  Suitable X-ray predictions
from CR-driven ISM models and galaxy formation models are needed to test the roles of galactic
fountains and of accreted extragalactic material in explaining the observed X-ray emission from the
Galactic halo.

\acknowledgements

We thank M.~R. Joung for helpful comments at the early stages of this work.
This research is based on observations obtained with \xmm, an ESA science mission with instruments
and contributions directly funded by ESA Member States and NASA.
We acknowledge use of the R software package \citep{R}.
D.B.H.\ acknowledges funding from NASA grants NNX08AJ47G and NNX13AF69G, awarded through the
Astrophysics Data Analysis Program.
R.L.S.\ acknowledges funding from NASA grant NNX13AJ80G, awarded through the Astrophysics Theory
Program.
K.K.\ acknowledges funding from KASI (Korea Astronomy and Space Science Institute) under the R\&D
program (Project No.\ 2013-1-600-01).
M.-M.M.L.\ acknowledges funding from NASA through the \chandra\ Theory Program under
grant TM011008X and the NSF under grant AST11-09395. Computations used here were performed on
machines of the NSF-sponsored XSEDE under allocation TG-MCA99S024.

\bibliography{references}

\begin{thebibliography}{55}
\expandafter\ifx\csname natexlab\endcsname\relax\def\natexlab#1{#1}\fi

\bibitem[{Agertz {et~al.}(2013)Agertz, Kravtsov, Leitner, \& Gnedin}]{agertz13}
Agertz, O., Kravtsov, A.~V., Leitner, S.~N., \& Gnedin, N.~Y. 2013, ApJ, 770,
  25

\bibitem[{Begelman \& McKee(1990)}]{begelman90b}
Begelman, M.~C., \& McKee, C.~F. 1990, ApJ, 358, 375

\bibitem[{Booth {et~al.}(2013)Booth, Agertz, Kravtsov, \& Gnedin}]{booth13}
Booth, C.~M., Agertz, O., Kravtsov, A.~V., \& Gnedin, N.~Y. 2013, ApJL, 777,
  L16

\bibitem[{Bregman \& Lloyd-Davies(2007)}]{bregman07}
Bregman, J.~N., \& Lloyd-Davies, E.~J. 2007, ApJ, 669, 990

\bibitem[{Breitschwerdt {et~al.}(1991)Breitschwerdt, McKenzie, \&
  V{\"o}lk}]{breitschwerdt91}
Breitschwerdt, D., McKenzie, J.~F., \& V{\"o}lk, H.~J. 1991, A\&A, 245, 79

\bibitem[{Breitschwerdt \& Schmutzler(1994)}]{breitschwerdt94}
Breitschwerdt, D., \& Schmutzler, T. 1994, Nature, 371, 774

\bibitem[{Carter {et~al.}(2011)Carter, Sembay, \& Read}]{carter11}
Carter, J.~A., Sembay, S., \& Read, A.~M. 2011, A\&A, 527, A115

\bibitem[{Chen {et~al.}(1997)Chen, Fabian, \& Gendreau}]{chen97}
Chen, L.-W., Fabian, A.~C., \& Gendreau, K.~C. 1997, MNRAS, 285, 449

\bibitem[{Crain {et~al.}(2010)Crain, McCarthy, Frenk, Theuns, \&
  Schaye}]{crain10}
Crain, R.~A., McCarthy, I.~G., Frenk, C.~S., Theuns, T., \& Schaye, J. 2010,
  MNRAS, 407, 1403

\bibitem[{Crain {et~al.}(2013)Crain, McCarthy, Schaye, Theuns, \&
  Frenk}]{crain13}
Crain, R.~A., McCarthy, I.~G., Schaye, J., Theuns, T., \& Frenk, C.~S. 2013,
  MNRAS, 432, 3005

\bibitem[{Cravens {et~al.}(2001)Cravens, Robertson, \& Snowden}]{cravens01}
Cravens, T.~E., Robertson, I.~P., \& Snowden, S.~L. 2001, JGR, 106 (A11), 24883

\bibitem[{de~Avillez \& Breitschwerdt(2004)}]{avillez04}
de~Avillez, M.~A., \& Breitschwerdt, D. 2004, A\&A, 425, 899

\bibitem[{de~Avillez \& Breitschwerdt(2012{\natexlab{a}})}]{avillez12a}
de~Avillez, M.~A., \& Breitschwerdt, D. 2012{\natexlab{a}}, ApJL, 756, L3

\bibitem[{de~Avillez \& Breitschwerdt(2012{\natexlab{b}})}]{avillez12b}
de~Avillez, M.~A., \& Breitschwerdt, D. 2012{\natexlab{b}}, ApJL, 761, L19

\bibitem[{Draine \& Giuliani(1984)}]{draine84}
Draine, B.~T., \& Giuliani, Jr., J.~L. 1984, ApJ, 281, 690

\bibitem[{Everett {et~al.}(2008)Everett, Zweibel, Benjamin, McCammon, Rocks, \&
  Gallagher}]{everett08}
Everett, J.~E., Zweibel, E.~G., Benjamin, R.~A., {et~al.} 2008, ApJ, 674, 258

\bibitem[{Ezoe {et~al.}(2010)Ezoe, Ebisawa, Yamasaki, Mitsuda, Yoshitake,
  Terada, Miyoshi, \& Fujimoto}]{ezoe10}
Ezoe, Y., Ebisawa, K., Yamasaki, N.~Y., {et~al.} 2010, PASJ, 62, 981

\bibitem[{Fang {et~al.}(2013)Fang, Bullock, \& Boylan-Kolchin}]{fang13}
Fang, T., Bullock, J., \& Boylan-Kolchin, M. 2013, ApJ, 762, 20

\bibitem[{Fang {et~al.}(2006)Fang, McKee, Canizares, \& Wolfire}]{fang06}
Fang, T., McKee, C.~F., Canizares, C.~R., \& Wolfire, M. 2006, ApJ, 644, 174

\bibitem[{Fujimoto {et~al.}(2007)Fujimoto, Mitsuda, McCammon, Takei, Bauer,
  Ishisaki, Porter, Yamaguchi, Hayashida, \& Yamasaki}]{fujimoto07}
Fujimoto, R., Mitsuda, K., McCammon, D., {et~al.} 2007, PASJ, 59, S133

\bibitem[{Gupta {et~al.}(2012)Gupta, Mathur, Krongold, Nicastro, \&
  Galeazzi}]{gupta12}
Gupta, A., Mathur, S., Krongold, Y., Nicastro, F., \& Galeazzi, M. 2012, ApJL,
  756, L8

\bibitem[{Hagihara {et~al.}(2010)Hagihara, Yao, Yamasaki, Mitsuda, Wang, Takei,
  Yoshino, \& McCammon}]{hagihara10}
Hagihara, T., Yao, Y., Yamasaki, N.~Y., {et~al.} 2010, PASJ, 62, 723

\bibitem[{Henley \& Shelton(2008)}]{henley08a}
Henley, D.~B., \& Shelton, R.~L. 2008, ApJ, 676, 335

\bibitem[{Henley \& Shelton(2012)}]{henley12b}
Henley, D.~B., \& Shelton, R.~L. 2012, ApJS, 202, 14

\bibitem[{Henley \& Shelton(2013)}]{henley13}
Henley, D.~B., \& Shelton, R.~L. 2013, ApJ, 773, 92 (HS13)

\bibitem[{Henley \& Shelton(2014)}]{henley14a}
Henley, D.~B., \& Shelton, R.~L. 2014, ApJ, 784, 54

\bibitem[{Henley {et~al.}(2014)Henley, Shelton, Cumbee, \& Stancil}]{henley14d}
Henley, D.~B., Shelton, R.~L., Cumbee, R.~S., \& Stancil, P.~C. 2014, ApJ, in
  press (arXiv:1411.5017)

\bibitem[{Henley {et~al.}(2010)Henley, Shelton, Kwak, Joung, \&
  Mac~Low}]{henley10b}
Henley, D.~B., Shelton, R.~L., Kwak, K., Joung, M.~R., \& Mac~Low, M.-M. 2010,
  ApJ, 723, 935 (H10)

\bibitem[{Hickox \& Markevitch(2006)}]{hickox06}
Hickox, R.~C., \& Markevitch, M. 2006, ApJ, 645, 95

\bibitem[{Hill {et~al.}(2012)Hill, Joung, Mac~Low, Benjamin, Haffner,
  Klingenberg, \& Waagan}]{hill12}
Hill, A.~S., Joung, M.~R., Mac~Low, M.-M., {et~al.} 2012, ApJ, 750, 104
 (H12; erratum 761, 189)
\bibitem[{Joung \& Mac~Low(2006)}]{joung06}
Joung, M.~K.~R., \& Mac~Low, M.-M. 2006, ApJ, 653, 1266
 (JM06)
\bibitem[{Joung {et~al.}(2012)Joung, Bryan, \& Putman}]{joung12a}
Joung, M.~R., Bryan, G.~L., \& Putman, M.~E. 2012, ApJ, 745, 148

\bibitem[{Joung {et~al.}(2009)Joung, Mac~Low, \& Bryan}]{joung09}
Joung, M.~R., Mac~Low, M.-M., \& Bryan, G.~L. 2009, ApJ, 704, 137

\bibitem[{Kennicutt(1998)}]{kennicutt98}
Kennicutt, Jr., R.~C. 1998, ApJ, 498, 541

\bibitem[{Koutroumpa {et~al.}(2007)Koutroumpa, Acero, Lallement, Ballet, \&
  Kharchenko}]{koutroumpa07}
Koutroumpa, D., Acero, F., Lallement, R., Ballet, J., \& Kharchenko, V. 2007,
  A\&A, 475, 901

\bibitem[{Kuntz \& Snowden(2000)}]{kuntz00}
Kuntz, K.~D., \& Snowden, S.~L. 2000, ApJ, 543, 195

\bibitem[{Lallement(2004)}]{lallement04b}
Lallement, R. 2004, A\&A, 422, 391

\bibitem[{Mac~Low {et~al.}(2012)Mac~Low, Hill, Joung, Waagan, Klingenberg,
  Wood, Benjamin, Federrath, \& Haffner}]{maclow12}
Mac~Low, M.-M., Hill, A.~S., Joung, M.~R., {et~al.} 2012, in ASP Conf. Ser.
  459, Numerical Modeling of Space Plasma Flows (ASTRONUM 2011), ed. N.~V.
  Pogorelov, J.~A. Font, E.~Audit, \& G.~P. Zank (San Francisco: ASP), 112

\bibitem[{Maller \& Bullock(2004)}]{maller04}
Maller, A.~H., \& Bullock, J.~S. 2004, MNRAS, 355, 694

\bibitem[{McKernan {et~al.}(2004)McKernan, Yaqoob, \& Reynolds}]{mckernan04}
McKernan, B., Yaqoob, T., \& Reynolds, C.~S. 2004, ApJ, 617, 232

\bibitem[{Moretti {et~al.}(2003)Moretti, Campana, Lazzati, \&
  Tagliaferri}]{moretti03}
Moretti, A., Campana, S., Lazzati, D., \& Tagliaferri, G. 2003, ApJ, 588, 696

\bibitem[{Nicastro {et~al.}(2002)Nicastro, Zezas, Drake, Elvis, Fiore,
  Fruscione, Marengo, Mathur, \& Bianchi}]{nicastro02}
Nicastro, F., Zezas, A., Drake, J., {et~al.} 2002, ApJ, 573, 157

\bibitem[{{R Development Core Team}(2008)}]{R}
{R Development Core Team}. 2008, {R: A Language and Environment for Statistical
  Computing}, R Foundation for Statistical Computing, Vienna, Austria

\bibitem[{Rasmussen {et~al.}(2003)Rasmussen, Kahn, \& Paerels}]{rasmussen03}
Rasmussen, A., Kahn, S.~M., \& Paerels, F. 2003, in The IGM/Galaxy Connection.
  The Distribution of Baryons at $z=0$, ed. J.~L. Rosenberg \& M.~E. Putman
  (Dordrecht: Kluwer), 109

\bibitem[{Raymond \& Smith(1977)}]{raymond77}
Raymond, J.~C., \& Smith, B.~W. 1977, ApJS, 35, 419

\bibitem[{Salem \& Bryan(2014)}]{salem14}
Salem, M., \& Bryan, G.~L. 2014, MNRAS, 437, 3312

\bibitem[{Shapiro \& Moore(1976)}]{shapiro76b}
Shapiro, P.~R., \& Moore, R.~T. 1976, ApJ, 207, 460

\bibitem[{Shelton(1998)}]{shelton98}
Shelton, R.~L. 1998, ApJ, 504, 785

\bibitem[{Smith {et~al.}(2007)Smith, Bautz, Edgar, Fujimoto, Hamaguchi, Hughes,
  Ishida, Kelley, Kilbourne, Kuntz, McCammon, Miller, Mitsuda, Mukai,
  Plucinsky, Porter, Snowden, Takei, Terada, Tsuboi, \& Yamasaki}]{smith07a}
Smith, R.~K., Bautz, M.~W., Edgar, R.~J., {et~al.} 2007, PASJ, 59, S141

\bibitem[{Snowden {et~al.}(2004)Snowden, Collier, \& Kuntz}]{snowden04}
Snowden, S.~L., Collier, M.~R., \& Kuntz, K.~D. 2004, ApJ, 610, 1182

\bibitem[{Snowden {et~al.}(2000)Snowden, Freyberg, Kuntz, \&
  Sanders}]{snowden00}
Snowden, S.~L., Freyberg, M.~J., Kuntz, K.~D., \& Sanders, W.~T. 2000, ApJS,
  128, 171

\bibitem[{Wargelin {et~al.}(2004)Wargelin, Markevitch, Juda, Kharchenko, Edgar,
  \& Dalgarno}]{wargelin04}
Wargelin, B.~J., Markevitch, M., Juda, M., {et~al.} 2004, ApJ, 607, 596

\bibitem[{Yao \& Wang(2007)}]{yao07a}
Yao, Y., \& Wang, Q.~D. 2007, ApJ, 658, 1088

\bibitem[{Yao {et~al.}(2009)Yao, Wang, Hagihara, Mitsuda, McCammon, \&
  Yamasaki}]{yao09}
Yao, Y., Wang, Q.~D., Hagihara, T., {et~al.} 2009, ApJ, 690, 143

\bibitem[{Yoshino {et~al.}(2009)Yoshino, Mitsuda, Yamasaki, Takei, Hagihara,
  Masui, Bauer, McCammon, Fujimoto, Wang, \& Yao}]{yoshino09}
Yoshino, T., Mitsuda, K., Yamasaki, N.~Y., {et~al.} 2009, PASJ, 61, 805

\end{thebibliography}

\end{document}